\documentclass[fleqn,usenatbib]{mnras}

\usepackage{newtxtext,newtxmath}

\usepackage[T1]{fontenc}

\DeclareRobustCommand{\VAN}[3]{#2}
\let\VANthebibliography\thebibliography
\def\thebibliography{\DeclareRobustCommand{\VAN}[3]{##3}\VANthebibliography}

\usepackage{graphicx}	
\usepackage{amsmath}	

\usepackage{textcomp}
\usepackage{graphicx}

\usepackage{enumerate}
\usepackage{epsfig}
\usepackage{caption}
\usepackage{subcaption}

\usepackage{setspace}
\usepackage{xcolor}
\usepackage{braket}
\usepackage{makecell}
\usepackage{multicol}
\usepackage{multirow}

\usepackage{ulem}

\title[Time variability properties of blazars]{RMS-Flux Relation and Disc-Jet Connection in Blazars in the Context of the Internal Shocks Model}

\author[A. Kundu et al.]{
Aritra Kundu,$^{1}$\thanks{E-mail: aritra8354@gmail.com}
Ritaban Chatterjee,$^{1}$
Kaustav Mitra$^{2,3}$
and Sripan Mondal$^{1}$
\\
$^{1}$Department of Physics,  Presidency University, 86/1 College Street, Kolkata 700073.\\
$^{2}$Department of Astronomy, 52 Hillhouse Avenue, Steinbach Hall, Yale University, New Haven, CT 06511, USA\\
$^{3}$Yale Center for Astronomy and Astrophysics, Yale University, PO Box 208120, New Haven, CT 06520-8120, USA\\
}

\date{Accepted XXX. Received YYY; in original form ZZZ}

\pubyear{2021}

\begin{document}
\label{firstpage}
\pagerange{\pageref{firstpage}--\pageref{lastpage}}
\maketitle

\begin{abstract}
Recent analysis of blazar variability has revealed a proportionality between the mean flux and the root mean squared (rms) fluctuations about the mean flux. Although such rms-flux relation has been previously observed in the accretion disc/corona variability of X-ray binaries and Seyfert galaxies, and has been extensively modelled, its emergence in the jet light curves of blazars calls for a revised theoretical understanding of this feature. In this work, we analyse the time variability properties of realistic multi-wavelength jet light curves, simulated in the context of a simplified version of the internal shocks model, particularly focusing on the rms-flux relation. These shocks accelerate the jet electrons to relativistic energies, which then cool radiatively via synchrotron and inverse-Compton processes. We find that the rms-flux relation may be consistently recovered in the cases, in which the shocks have different amplitudes based on the speed of the colliding blobs generating them as opposed to all shocks having the same amplitude. We observe that the slope of the rms-flux relation depends on the wavelength at which the variability is observed and the energy distribution of the electron population. We find that the accretion disc and the jet variability are anti-correlated, with the latter lagging that of the disc. Our results provide crucial constraints on the physical properties of the jet, and the mode of connection through which the accretion disc and jet may be related.

\end{abstract}




\section{Introduction}

Active Galactic Nuclei (AGN) are the compact regions in the central part of certain galaxies in which a supermassive black hole (SMBH) actively accretes matter from its surroundings in the form of a planar inflow. In addition, some AGNs show the presence of two relativistic jets emanating perpendicular to the plane of accretion \citep[see][for recent detailed reviews]{Boettcher&etalbook2012,Antonucci2012,Krawczynski&etal2013,Padovani&etal2017,Romero&etal2017, Blandford&etal2019,Hovatta&etal2019}. A class of AGNs, called blazars, have their jets pointed towards us, making an angle of $10^{\circ}$ or less with our line of sight \citep{Urry&Padovani1995,Netzer2015review}. Due to this, the emission from the jet is relativistically beamed and hence the observed flux of the blazar is dominated by the jet compared to the emission from the other parts of the AGN such as the accretion disc, broad line region and torus. The formation process and ongoing physical mechanisms of these jets are not well understood. Blazars are one of the best class of objects to study jet physics.

One of the key characteristics of blazars is the high amplitude variability in their observed flux, across a wide range of the electromagnetic spectrum, sometimes from radio to TeV $\gamma$-rays with timescales ranging from years to even sub-hour \citep{Ulrich&etal1997,Marscher2016}. Analysing the variability is an important aspect for understanding the mechanisms involved in the jet. One way of analysing the jet light curves is by looking at the Fourier frequency dependent properties of the variability. But recently, the temporal properties of the jet light curves in real space, have been used to constrain the underlying stochastic processes in a manner complementary to that of the frequency domain. One such temporal property is the rms-flux relation, which was initially found in the light curves of accreting objects like X-ray binaries (XRBs)  \citep{Uttley&McHardy2001, Gleissner&etal2004, Uttley&etal2005, Heil&etal2012, Dobrotka&Ness2015, Gandhi2009}, ultraluminous X-ray sources \citep{Heil&Vaughan2010}, accreting white dwarfs \citep{Scaringi&etal2012}, and in the X-ray variability of Seyfert 1 galaxies \citep{Gaskell2004, Alston&etal2019}. It is observed that the amplitude of rms variability increases linearly with the mean flux, implying that the source is more variable when it is brighter. The presence of an rms-flux relation in an observed time series implies a connection between the short and long timescale variations and points toward an underlying multiplicative stochastic process \citep{Uttley&etal2005, Vaughan&etal2003}. Although one can use power spectral analysis to constrain different variability models, a linear rms-flux relation was observed in the black hole X-ray binary (BHXRB) Cyg X-1 in all of its known spectral states, independent of the power spectral density (PSD) shape \citep{Gleissner&etal2004}, and thereby suggesting that the rms-flux relation is probably more robust in constraining the variability models than the PSD.

The rms-flux relation in the above objects is observed in the X-ray variability, which mainly originates from the accretion disc-corona region. The leading scenario to explain the rms-flux relation in these objects, is the so-called ``propagating fluctuation model,'' in which  short-timescale fluctuations in the inner disc are superimposed on the longer-timescale fluctuations in the outer disc while matter falls inward giving rise to a multiplicative nature of the variability \citep{Lyubarskii1997, Arevalo&Uttley2006}. But in recent years, the relation has been found in the jet light curves of blazars. \cite{Giebels&Degrange2009} first observed a linear rms-flux relation and log-normal flux distribution in the blazar BL Lac, while \cite{Edelson&etal2013} found a non-linear rms-flux relation by analysing the optical light curves of the BL Lac object W2R1926+42. The possible connection between log-normal distribution of the flux and rms-flux relation have also been found in some Fermi selected blazars \citep{Kushwaha&etal2017, Bhatta&Dhital2020}, though it might not be a robust connection. Given the geometry and emission mechanisms involved in the disc and the jet are different, the presence of rms-flux relation in jet light curves may be an indication of a possible mode of the disc-jet connection in blazars \citep{Giebels&Degrange2009}.

Understanding the relation between the dynamics of the disc and the jet is crucial to put constraints on the theories of jet launching and collimation. In BHXRBs, it has been observed that dips in the X-ray emission, presumably due to sudden collapse of the inner disc into the BH, is followed by ejection of blobs in the jet \citep{Fender&etal2004, Fender&etal2009, Remillard&etal2006}. It is generally assumed that the jet is launched from a region close to the accretion disc. Hence, some connection between the disc and the jet is expected as a natural consequence. Similar correlation between the dips in the X-ray emission from the disc-corona region and the outbursts in the jet have been observed in AGNs \citep{Marscher&etal2002, Chatterjee&etal2009, Chatterjee&etal2011}. \cite{Mukherjee&etal2019} analysed the connection between the break time-scales of the PSDs of the jet and the disc, by assuming them to be related to the interval of large outbursts in the jet and the viscous time-scales in the disc respectively, but did not identify any simple relation between those. \cite{Casella&etal2010} found a correlation between the infrared and X-ray light curves of BHXRBs, with the infrared variation lagging behind that of the X-rays. The rms-flux relation provides another window for probing the disc-jet connection in blazars. Though numerical simulations of the accretion disc variability \citep{Cowperthwaite&Reynolds2014}, along with the effects of magnetohydrodynamics \citep{Hogg&Reynolds2016} are starting to reproduce the rms-flux relation and the observed flux distributions, a clear physical model for the relation is yet to be proposed in the case of the blazar light curves.

When material from the inner regions of the accretion disc plunges into the black hole, some part of it gets ejected down the jet as relativistic blobs of plasma \citep{Merloni&etal2003, Falcke&etal2004, Malzac&etal2004, Gultekin&etal2009}. In the internal shocks model \citep{Spada&etal2001, Jamil&etal2010, Malzac2013, Malzac2014}, those blobs have a distribution of bulk Lorentz factors (LF) and thus can collide due to the difference in their relative velocities, leading to formation of shock fronts in the jet. When a shock front passes through a region, particles in that location, e.g., electrons get energized. Those electrons then cool via radiative mechanisms like synchrotron emission and inverse-Compton (IC) scattering \citep{Jones&Odell1977}, and the whole process leads to a non-thermal flaring event. 

In this work, we simulate blazar light curves using the above scenario. We study the possible presence of rms-flux relation in the simulated light curves and investigate how the nature of the relation, if present, changes as we vary the distribution of the bulk Lorentz factors, amplitude of the shocks produced by the collision of the blobs moving down the jet, location of the synchrotron peak of the blazar SED as determined by the maximum Lorentz factor of the electron population energized by the shock, and the waveband of the analysed light curve. We search for conditions that give rise to the rms-flux relation in order to put constraints on the parameters of the above model.

In section 2, we describe the numerical modelling of the jet structure, the radiative mechanisms and the internal shocks model used in our simulation. We show the results of our simulation and analysis in section 3 and in section 4 we present the discussion and summary.

\section{Numerical Modelling}

We use a computer program to simulate the production of shock fronts following the internal shocks model as described in the previous section. We consider a cylindrical emission region in the jet and divide it into $80$ cells along its length, each of which has its individual electron population and magnetic field. We pass the shock fronts through the emission region and calculate the synchrotron and inverse-Compton emission in each cell, at different frequencies and at different instances of time to generate the light curves. The magnetic field in the emission regions has a gradient and its value at the upstream and downstream ends are kept as $1$ $G$ and $0.3$ $G$, respectively \citep{Majumder&etal2019}.

\subsection{Synchrotron emission}

The double-humped spectral energy distribution (SED) of blazars is due to the non-thermal emission processes in the jet \citep{Abdo&etal2010, Abdo&etal2011}. The low frequency hump corresponds to synchrotron emission and mostly contributes to the continuum emission from radio to optical \citep{Urry&Mushotzky1982, Impey&etal2000, Marscher1998}. The electrons follow a power-law distribution of energy and the total synchrotron emissivity $j_S(\nu)$ as a function of the frequency of the emitted photons $\nu$ is given by \citep{Rybicki&Lightman1979} 

\begin{equation}
    j_S(\nu) = \frac{\sqrt{3} e^3}{m_e c^2}B sin \psi\int_{\gamma_{min}}^{\gamma_{max}} N(\gamma, t) d\gamma \,x \int_x ^{\infty} K_{5/3}(\xi) d\xi,
    \label{eqn: 1}
\end{equation}
where $N(\gamma, t)$ is the time variable electron energy distribution and $\gamma_{min}$ and $\gamma_{max}$ are the minimum and maximum Lorentz factors of the electron energy distribution, respectively. $x$ = $\nu/\nu_c$, where $\nu_c$ is the critical frequency and is given by $\nu_c = k_1 \gamma^2$, and $k_1 = 4.2 \times 10^6 B$, where $B$ is the magnetic field in $G$, $\psi$ is the angle between $\vec{B}$ and the line of sight and $K_{5/3}$ is the modified Bessel function of order 5/3. Depending on the value of $\gamma_{max}$ the synchrotron hump may extend into the UV to X-ray energies. 

\subsection{Inverse-Compton scattering}

The high frequency hump in blazar SEDs correspond to inverse-Compton scattering, in which low energy synchrotron photons, which are produced in the jet itself, are up-scattered in a process termed synchrotron self-Compton (SSC) \citep{Maraschi&etal1992, Chiang&Bottcher2002}, or the photons from external regions, such as the BLR and torus, are incident on the jet and get up-scattered via the so-called ``external Compton (EC)'' process \citep{Sikora&etal1994, Blazejowski&etal2000, Dermer&etal2009}.  

\medbreak
The IC emission coefficient $j_{IC}(\nu)$ is given by \citep{Rybicki&Lightman1979}, 

\begin{equation}
    j_{IC}(\nu) = \int_{\nu_i} \int_{\gamma} \frac{\nu_f}{\nu_i} j_{\nu}(\nu_i) R \sigma(\nu_i, \nu_f, \gamma) N(\gamma, t) d\gamma d\nu_i,
    \label{eqn: 2}
\end{equation}

where $\nu_i$ and $\nu_f$ are the frequencies of the photons before and after the scattering, $R$ is the length scale associated with the emission region, and $j_\nu(\nu_i)$ is the emissivity of the incident photon field which may be from within or outside the jet.

\medbreak
The inverse-Compton scattering cross-section $\sigma$ depends on the incident frequency $\nu_i$, scattered frequency $\nu_f$ and the Lorentz factor $\gamma$. It is given by \citep{Blumenthal&Gould1970},

\begin{equation}
    \sigma(\nu_i, \nu_f, \gamma) = \frac{3}{32} \frac{\sigma_{KN}}{\nu_i \gamma^2} \big[8 + 2x - x^2 + 4x\, ln(x/4) \big],
    \label{eqn: 3}
\end{equation}

where $x = \nu_f/(\nu_i \gamma^2$) and $\sigma_{KN}$ is the Klein-Nishina cross-section \citep{Beckmann&Shrader2012}.

\medbreak
We model the SSC process in a particular cell by adding the contribution of the seed photons from all the other cells, while giving weightage according to their inverse-squared distance from the former, and taking into account the light-propagation delay of the synchrotron photons \citep{Majumder&etal2019,royetal2019,graffetal2008}. For the EC process, we take the contributions of the incident radiation field from the broad line region (BLR) and the torus. We model the energy density of the BLR and torus, respectively, as follows \citep{Hayashida&etal2012},

\begin{equation}
    u_{BLR}(r) = \frac{\epsilon_{BLR} \Gamma_{jet}^2 L_D}{3\pi r_{BLR}^2 c [1 + (r/r_{BLR})^{\beta_{BLR}}]}
    \label{eqn: 4}
\end{equation}

\begin{equation}
    u_{torus}(r) = \frac{\epsilon_{torus} \Gamma_{jet}^2 L_D}{3\pi r_{torus}^2 c [1 + (r/r_{torus})^{\beta_{torus}}]},
    \label{eqn: 5}
\end{equation}

where $r$ is the distance of the cell from the central engine. The bulk Lorentz factor of the jet $\Gamma_{jet}$ is taken to be $10$, $c$ is the speed of light, $L_D$ is the accretion disc luminosity, $\epsilon_{BLR}$ $=$ $0.1$ and $\epsilon_{torus}$ $=$ $0.01$ are the fraction of disc luminosity reprocessed into emission lines and into hot dust radiation, respectively. Here $r_{BLR}$ is the distance of the BLR from the central engine and is taken to be $0.9$ pc, while $r_{torus}$ is the distance of the torus from the central engine and is taken to be $10$ pc. We take $\beta_{BLR}$ $=$ $3$ and $\beta_{torus}$ $=$ $4$ \citep{Hayashida&etal2012}.  We consider the radiation from the BLR and torus to be in the ultraviolet (UV; $10^{15.6}$ Hz)  and infrared (IR; $10^{12}$ Hz) waveband respectively, and then add them to the incident emissivity accordingly.

\medbreak
The time variable electron energy distribution is modelled as follows
\begin{equation}
    N(\gamma, t) = N_0 \gamma^{-s} (1 - \gamma k_2 t),
    \label{eqn: 6}
\end{equation}

$N_0$ being the total number of electrons, which acts as a normalization factor in our case. Here $s = 2.5$ and $k_2 = \frac{4}{3}\frac{\sigma_T c}{m_e c^2}(U_B + U_R)$, where $\sigma_T$ is the Thomson cross-section, $U_B$ is the magnetic energy density and $U_R$ is the radiation energy density \citep{Karsashev1962, Ciprini2010}. The term $(1 - \gamma k_2 t)$ incorporates the radiative cooling of the electron distribution due to which $\gamma$ changes over a certain cooling timescale $t$, which we have taken to be the time resolution of our simulation. The change in $\gamma$ due to radiative cooling follows the equation $d\gamma / dt$ $=$ $-k_2 \gamma^2$.

Our simulation depends on a number of parameters of the jet which we provide as input. We keep certain parameters constant in every run of the simulation. These constant input parameters are listed in Table \ref{table:1}. 

\medbreak
\begin{table}
    \centering
    \caption{Input parameters of the jet which are kept constant in each run of the simulation.}
    \setlength{\tabcolsep}{1pt}
    \renewcommand{\arraystretch}{1.5}
    
    \begin{tabular}{lc}
        \hline\hline
        Parameters & Values\\
        \hline
        Minimum Lorentz factor of the electrons, $\gamma_{min}$   & $30$\\
        Magnetic field at the upstream end of the emission region, $B_i$ (in G) & 1.0\\
        Magnetic field at the downstream end of the emission region, $B_f$ (in G) & 0.3\\
        Accretion disc luminosity, $L_D$ (in $erg$ $s^{-1}$) & $10^{45}$\\
        Power of the electron distribution, $s$ & 2.5\\
        \hline\hline
    \end{tabular}
    
    \label{table:1}
    
\end{table}

\subsection{Internal shocks model and generation of bulk Lorentz factors}

\begin{figure*}
    \centering
        \includegraphics[width=\textwidth, height=0.3\textwidth]{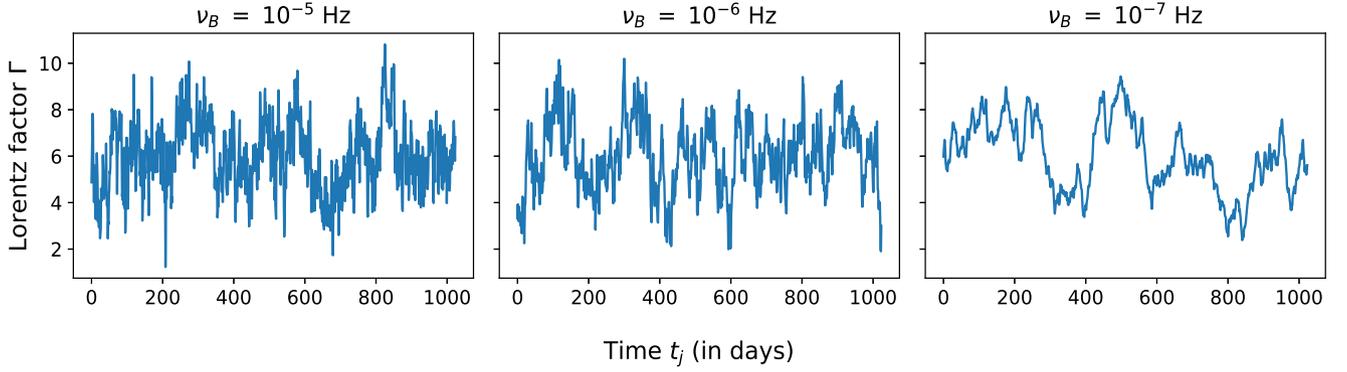}
        \caption{Lorentz factor variation for three different break frequencies of the generating PSD. \textit{Left:} $\nu_B$ $=$ $10^{-5}$ Hz, \textit{Middle:} $\nu_B$ $=$ $10^{-6}$ Hz and \textit{Right:} $\nu_B$ $=$ $10^{-7}$ Hz. The values are scaled to be within 1 to 11. The X-axis denotes the injection time $t_j$ (in days) of the blobs (see text for more details).}
        \label{fig:1}
\end{figure*}

In the internal shocks framework, the jet is modelled using discrete blobs of plasma, which have a distribution of bulk Lorentz factors, i.e., different velocities.  A blob with a larger Lorentz factor injected at a later time, can catch up and collide inelastically \citep{Lazzati&etal1999} with a preceding blob with a smaller Lorentz factor, thereby leading to the formation of shock fronts which travel down the jet and energizes the electrons. We model the jet using a simple version of the internal shocks model by mainly focusing on the physics after the formation of the shock fronts , e.g., the shock injection times in the emission region and the amplitude of the shocks. The disc emission variability of AGNs is sometimes characterized by a PSD which has a shape of a broken or bending power-law \citep{McHardy&etal2004, Chatterjee&etal2009}. \cite{Drappeau&etal2015} showed that the observed radio-infrared SED of the BHXRB GX 339-4 can be reproduced, if the PSD of the Lorentz factor fluctuations is same as the simultaneously observed X-ray PSD. Following them, we assume that the Lorentz factor fluctuations trace the disc variability \citep{Malzac&etal2018} and generate the bulk Lorentz factor distribution from a broken power-law PSD, using the \cite{Timmer&Koenig1995} algorithm. The PSD is characterized by \citep{McHardy&etal2004, Chatterjee&etal2009}

\begin{equation}
    P(\nu) = A \nu^{-\alpha_L} \Bigg[ 1 + \Bigg(\frac{\nu}{\nu_B}\Bigg)^{(\alpha_H - \alpha_L)} \Bigg]^{-1},
    \label{eqn: 7}
\end{equation}

where $A$ is the normalization constant, $\nu_B$ is the break frequency, $\alpha_H$ and $\alpha_L$ are the high- and low-frequency slope of the power law, respectively. 

We use $\alpha_H$ $=$ $2.5$ and $\alpha_L$ $=$ $1.0$, which are the values found in the X-ray PSD of BHXRBs and Seyfert galaxies \citep[e.g.,][]{Remillard&etal2006, mchardy2006}. Using $L_{\rm bol}$ = $\rm 10^{-3}-10^{-1}L_{edd}$, $M_{\rm BH} = 10^7 - 10^9$ M$_{\sun}$, and the best-fit values and uncertainties in the $\rm T_B$---$\rm M_{\rm BH}$---$\rm L_{\rm bol}$ relation proposed by \citet{mchardy2006}, expected break frequency ($\rm T_B$) is in the range  $10^{-5} - 10^{-7}$ Hz. Therefore, we consider three cases of the break frequency, $\nu_B$ $=$ $10^{-5}$ Hz, $10^{-6}$ Hz and $10^{-7}$ Hz, to generate three sets of Lorentz factor variation curves. The range of values of the break frequency are consistent with those found in the PSDs of observed blazars \citep{Kataoka&etal2001, McHardy2008, Isobe&etal2015, Chatterjee&etal2018}. Fig. \ref{fig:1} shows the Lorentz factor series for the three different break frequencies.

\medbreak
The time series of the Lorentz factors are normalized in such a way that keeps their values within 1 to 11, which is the approximate range of bulk Lorentz factor values of blazar jets that we are using. We keep the time interval ($\delta t_{inj}$) between the ejection of successive blobs to be constant at $10$ days \citep{Malzac&etal2018}. If the bulk Lorentz factor of a blob is $\Gamma$, then its velocity is given by $\beta = \sqrt{1 - \frac{1}{\Gamma^2}}$. The relative velocity between two blobs may be calculated using relativistic velocity addition formula,  

\begin{equation}
    v_{rel} = \frac{\beta_j - \beta_{j-1}}{1 - \beta_j \beta_{j-1}},
    \label{eqn: 8}
\end{equation}

where $\beta_j$ and $\beta_{j-1}$ are the velocities of the $j^{th}$ and $j-1^{th}$ blob respectively. 

\medbreak
 The time of collision of two blobs, which is also the time of shock injection, is then,

\begin{equation}
    t_{coll} = \frac{\beta_{j-1} \delta t_{inj}}{v_{rel}} + t_j,
    \label{eqn: 9}
\end{equation}

where $t_j$ is the injection time of the $j^{th}$ blob. 

\medbreak
Following the internal shocks model in \cite{Jamil&etal2010}, we calculate the internal energies of the merged blobs, which are given by,

\begin{equation}
    E_{int} = \Gamma_j + \Gamma_{j-1} - 2\Gamma_{merged},
    \label{eqn: 10}
\end{equation}

where $\Gamma_{merged}$ is the Lorentz factor of the merged blob after collision. It is calculated from the relative velocity of the merged blob, which is given by

\begin{equation}
    \beta_{merged} = \frac{\Gamma_j \beta_j + \Gamma_{j-1} \beta_{j-1}}{\Gamma_j + \Gamma_{j-1}}.
    \label{eqn: 11}
\end{equation}

The internal energies of the merged blobs serve as a proxy for the amplitude of the shocks in our simulation. If the effect of internal energies is not taken into account, then it is assumed that the shock amplitudes are the same. In our calculations, we assume same mass for all the blobs.

\section{Results}

\begin{figure*}
    \begin{subfigure}{0.49\textwidth}
        \centering
        \includegraphics[width=\textwidth, height=0.8\textwidth]{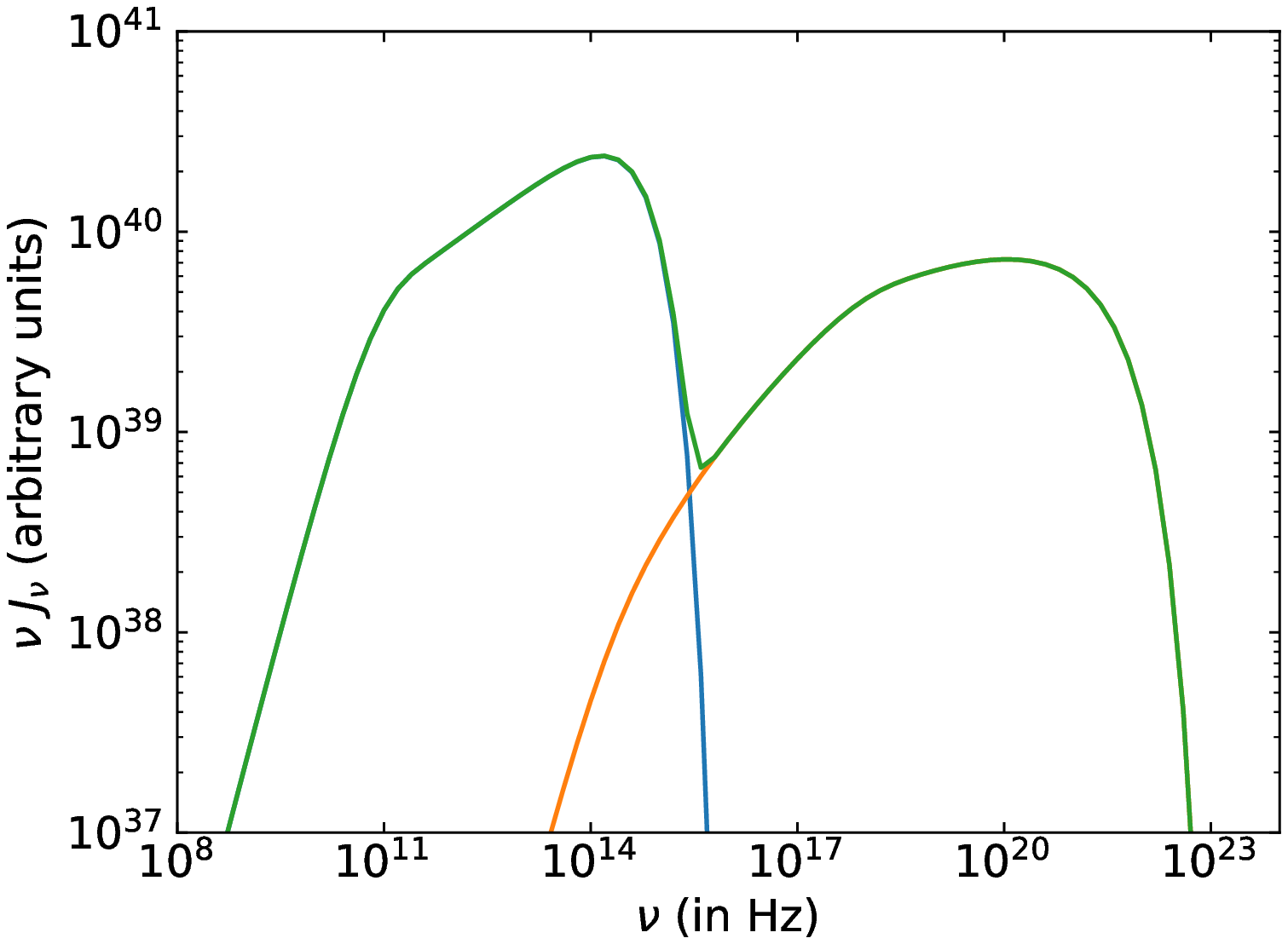}
        \caption{}
        \label{fig:2a}
    \end{subfigure}
    \hfill%
    \begin{subfigure}{0.49\textwidth}
        \centering
        \includegraphics[width=\textwidth, height=0.8\textwidth]{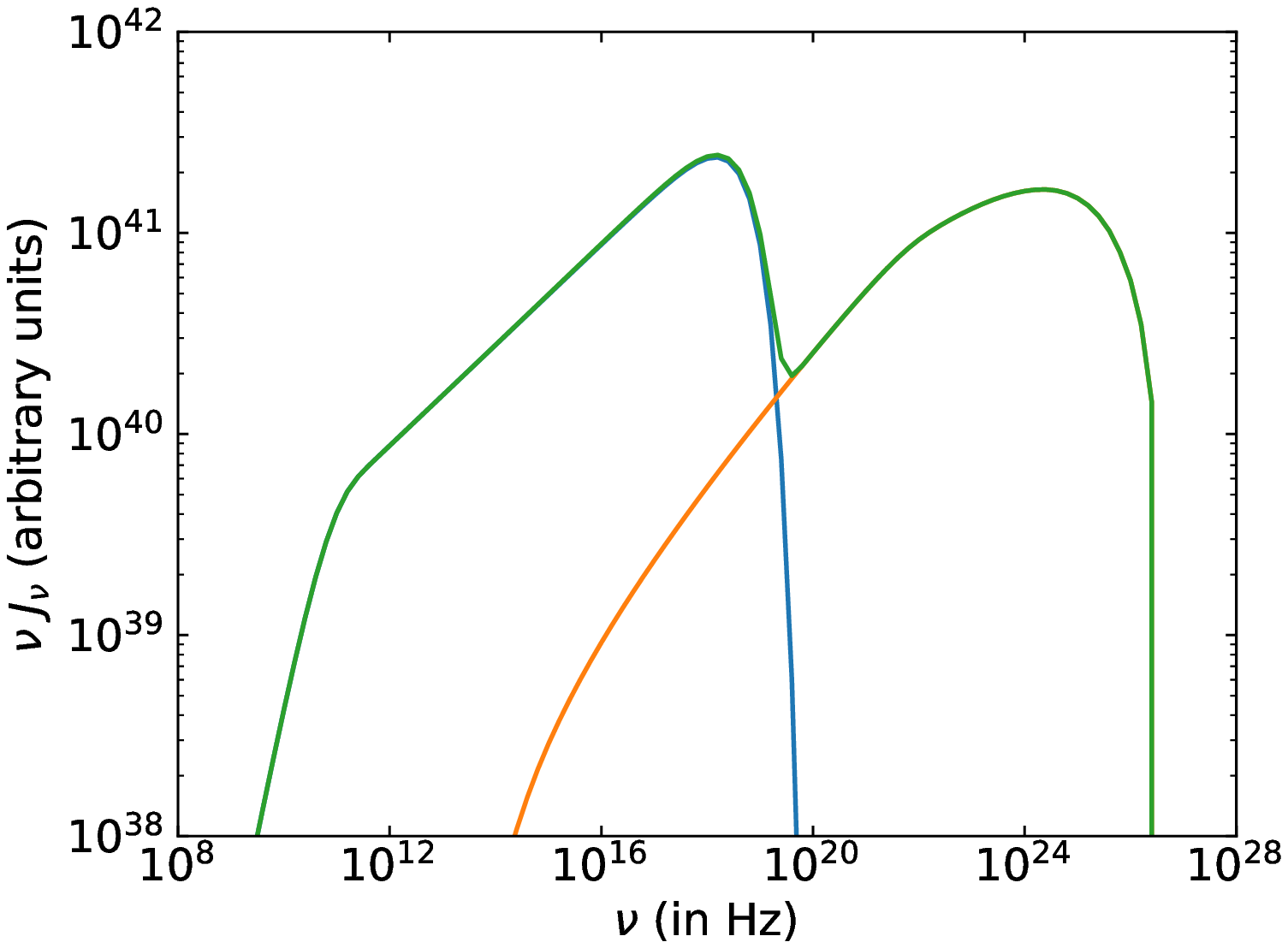}
        \caption{}
        \label{fig:2b}
    \end{subfigure}
    \caption{Snapshot SEDs from the simulation, where (a) $\gamma_{max}$ $=$ $3 \times 10^{3}$ and (b) $\gamma_{max}$ $=$ $3 \times 10^{5}$. The plots were generated by running the simulation for a single shock and then adding the contribution from all the cells at a given time. The simulated SEDs were transformed from the jet frame to the observer's frame using the Doppler factor $\delta$ ($\sim$10). The similarity of the simulated SEDs with the observed ones provide confirmation that our simulation includes the physics that is relevant in producing the broad features of blazar SEDs.}
     \label{fig:2}
\end{figure*}

We run our simulation and generate multi-wavelength light curves of the jet starting with the above three PSD models of the Lorentz factor distribution. In each case, we consider two values of the maximum Lorentz factor of the electron population, $\gamma_{max}$ $=$ $3 \times 10^{3}$ and $3 \times 10^{5}$ such that the resultant SEDs are approximately similar to those of low-synchrotron peaked (LSP) and high-synchrotron peaked (HSP) blazars, respectively. All of the calculations are done in the jet frame since we do not fit the model light curves with observed data and we intend to avoid introducing uncertainties due to the Doppler factor, the value of which varies over a wide range for the blazar population. We simulate the light curves for $\sim$1100 days. This length was chosen in order to obtain the longest possible light curve within a reasonable computational time and a short time resolution. We consider two scenarios: One in which the amplitude of the resultant shocks are related to the internal energies of the merged blobs, and hence different in different events, and another, in which the said amplitudes are related to the mean of all the internal energies, and therefore equal in all cases. In our model, the amplitude of the shock is directly related to the number or density of the electrons that are energized by them and hence it determines the normalization of the energised electron energy distribution. We use these two scenarios for both the values of $\gamma_{max}$ and also for each of the PSD models of the Lorentz factor distribution. We show our results at infrared, X-ray and $\gamma$-ray wavebands in order to sample parts of the SED dominated by synchrotron, SSC, and EC processes. Finally, we perform a cross-correlation analysis between the disc light curves (Lorentz factor series) and the jet light curves.

\subsection{Spectral Energy Distribution (SED)}

The SED constructed from the broadband blazar emission simulated using the above model with $\gamma_{max}$ $=$ $3 \times 10^{3}$ and $3 \times 10^{5}$ are shown in Figures \ref{fig:2a} and \ref{fig:2b}, respectively. They are constructed by passing a single shock in the jet and then adding the emission from all the cells at a given time. We transformed the SEDs from the jet frame to the observer's frame by using the Doppler factor $\delta (\sim$10), which resulted in a shift of the frequency axis toward higher values by an order of magnitude. The y-axis is not affected because it is in arbitrary units. The SEDs clearly show the standard double humped feature and when $\gamma_{max}$ is more, the synchrotron and IC peaks move to higher frequencies, as expected.

\subsection{Light curves}

\begin{figure*}
    \begin{subfigure}{\textwidth}
        \centering
        \includegraphics[width=0.95\textwidth, height=0.59\textwidth]{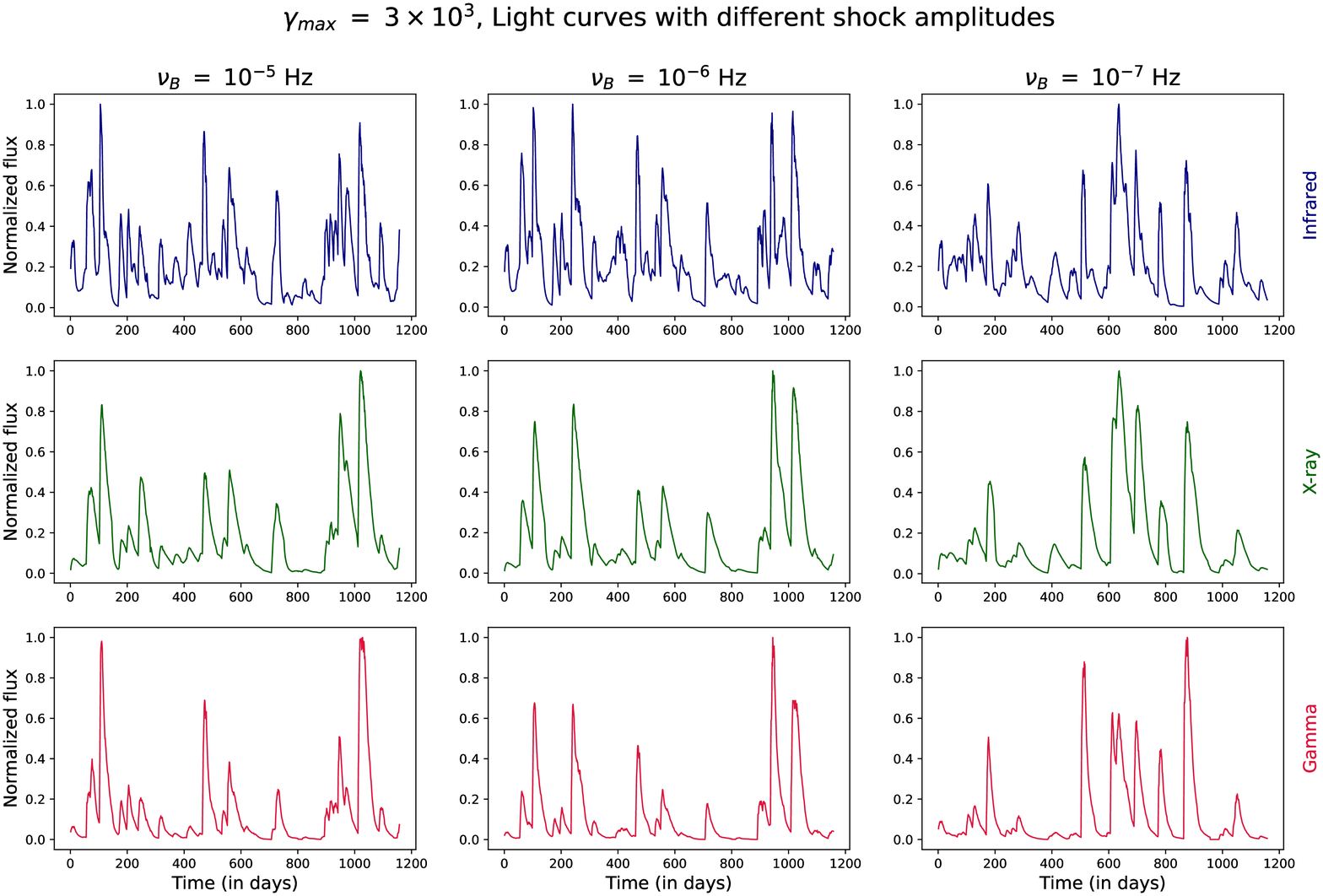}
        \caption{}
        \label{fig:3a}
    \end{subfigure}
    \hfill%
    \begin{subfigure}{\textwidth}
        \centering
        \includegraphics[width=0.95\textwidth, height=0.59\textwidth]{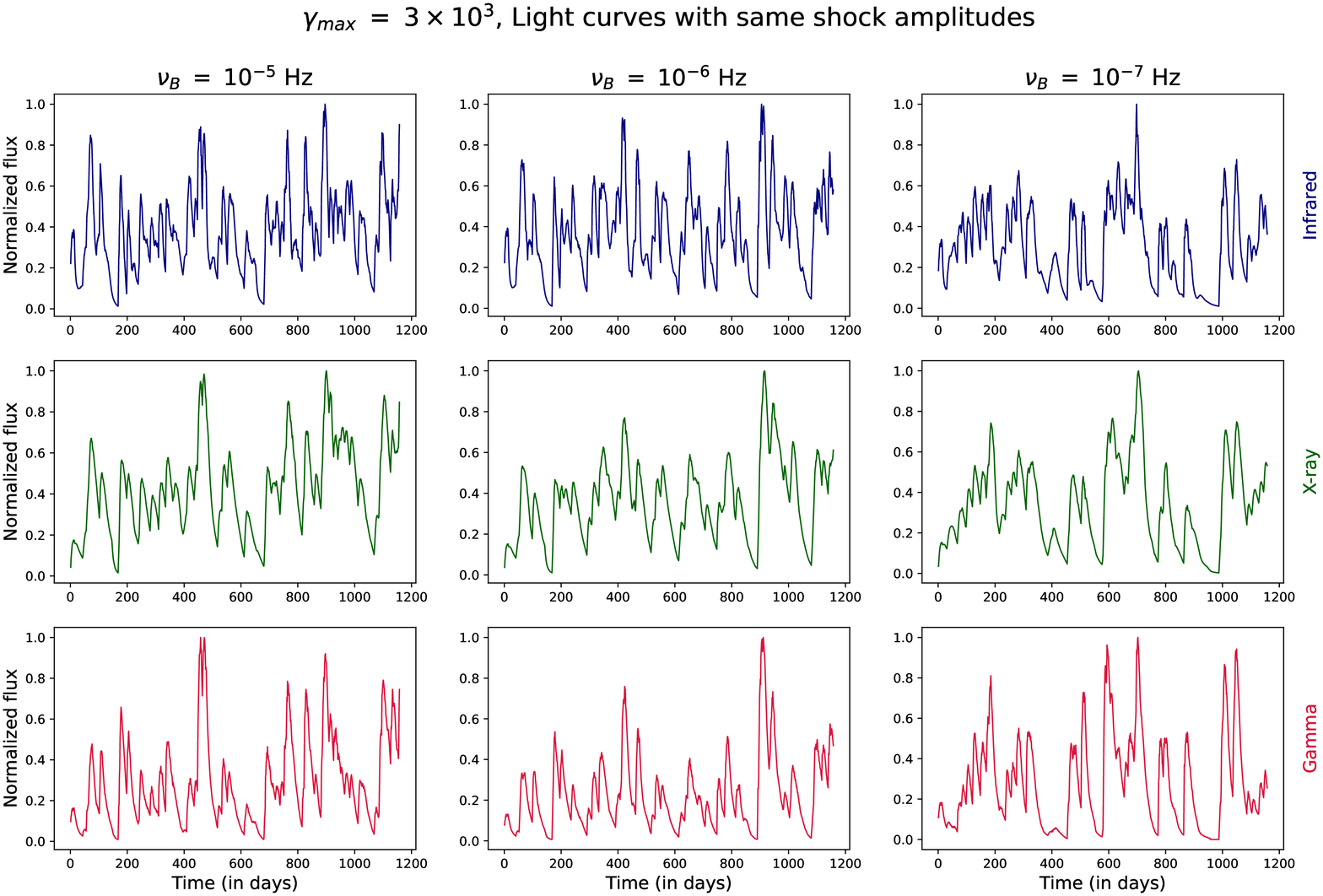}
        \caption{}
        \label{fig:3b}
    \end{subfigure}
    \caption{Multi-wavelength light curves simulated in this work for $\gamma_{max}$ $=$ $3 \times 10^{3}$ and three different break-frequencies of the PSD of the Lorentz factor fluctuations. Blue, green, and red solid lines denote infrared ($10^{12.0}$ Hz),  X-ray ($10^{18}$ Hz) and  $\gamma$-ray ($10^{20.6}$ Hz) light curves, respectively. (a) Different shock amplitudes are considered, (b) Same amplitude for all the shocks are considered.}
    \label{fig:3}
\end{figure*}

\begin{figure*}
    \begin{subfigure}{\textwidth}
        \centering
        \includegraphics[width=0.95\textwidth, height=0.59\textwidth]{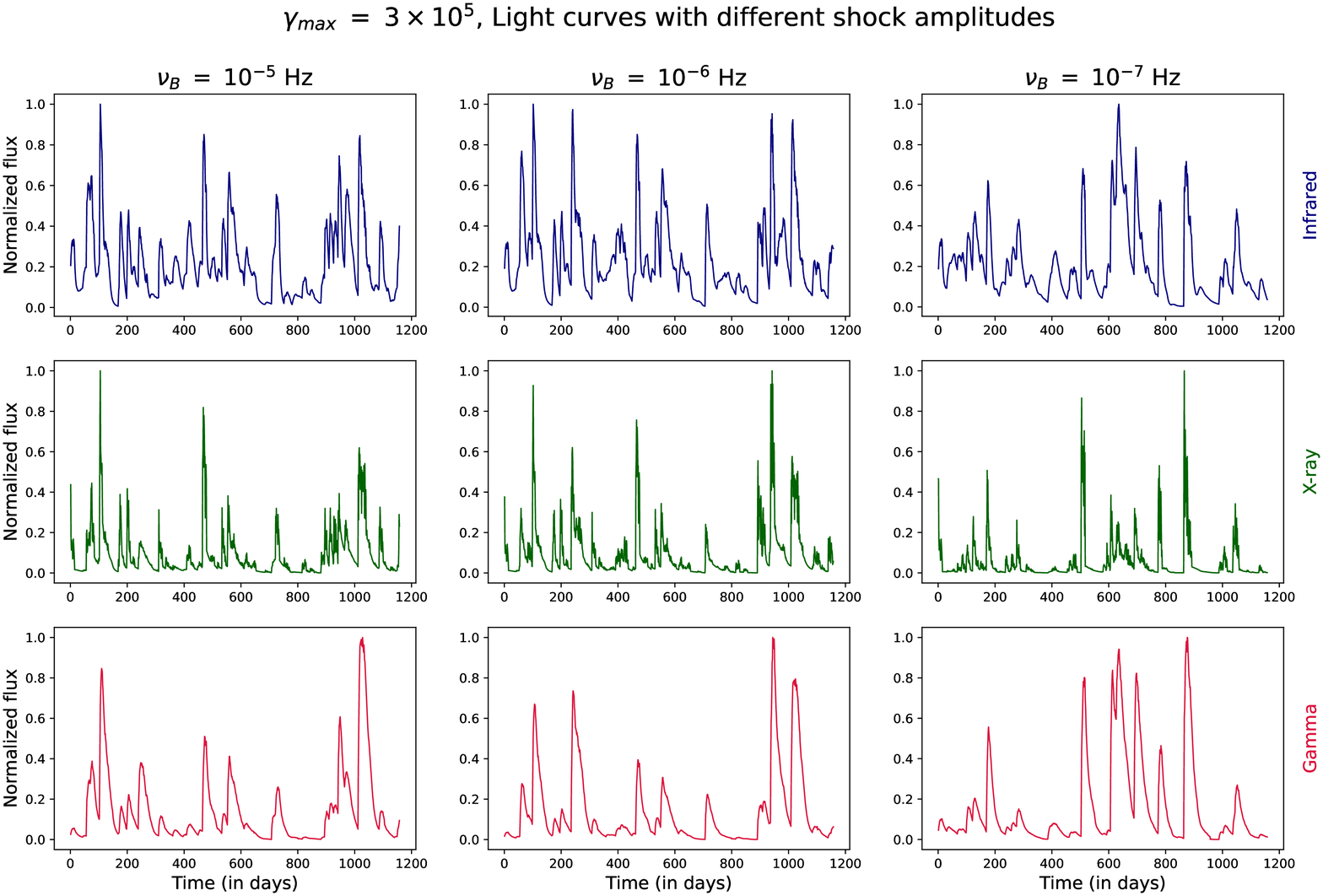}
        \caption{}
        \label{fig:4a}
    \end{subfigure}
    \hfill%
    \begin{subfigure}{\textwidth}
        \centering
        \includegraphics[width=0.95\textwidth, height=0.59\textwidth]{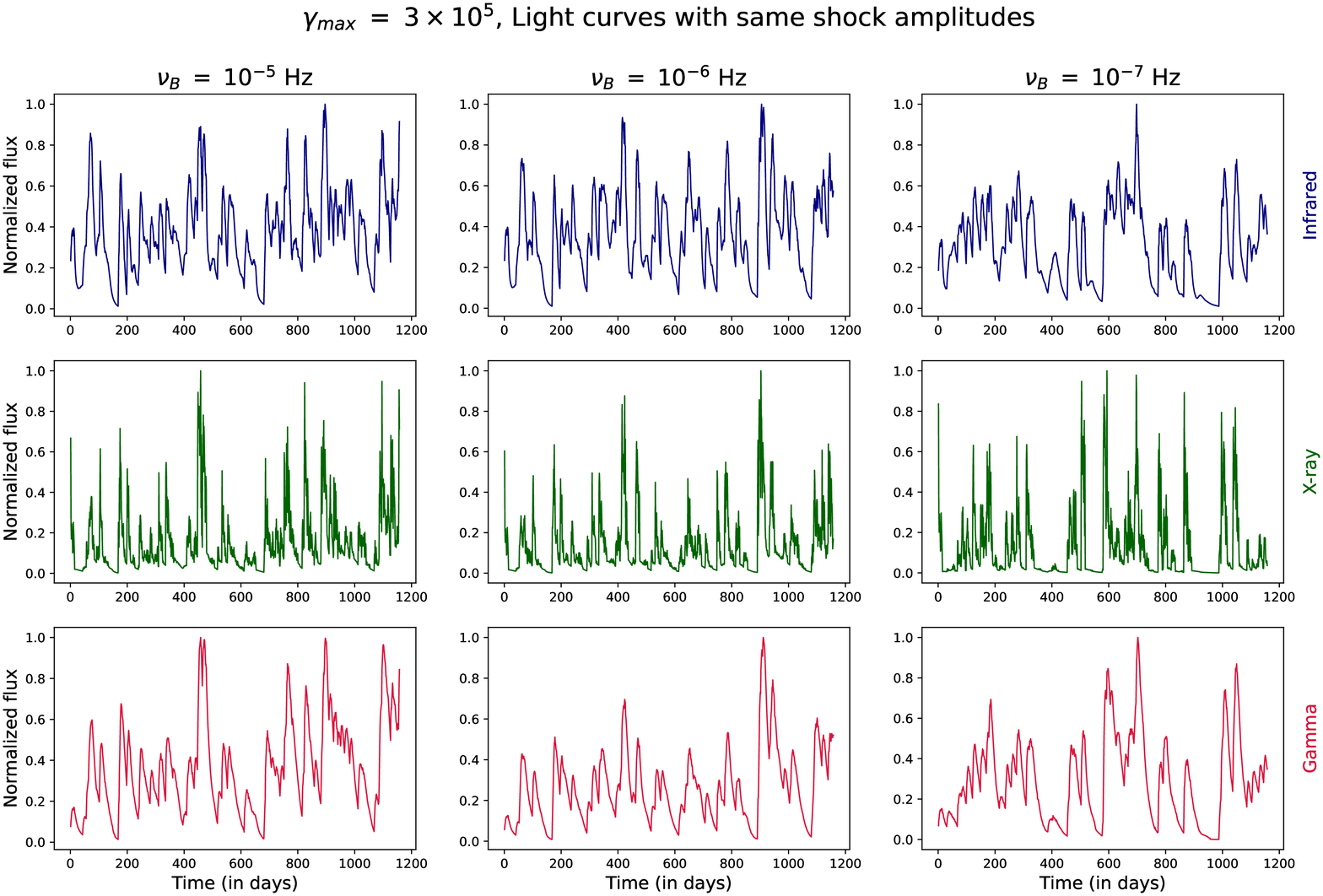}
        \caption{}
        \label{fig:4b}
    \end{subfigure}
    \caption{Same as Figure 3 for $\gamma_{max}$ $=$ $3 \times 10^{5}$. Here, the X-rays are produced by the high energy end of the electron distribution by the synchrotron process and hence decay faster than their $\gamma$-ray counterparts. See text for more details.}
    \label{fig:4}
\end{figure*}

We show the light curves for $\gamma_{max}$ $=$ $3 \times 10^{3}$ and $\gamma_{max}$ $=$ $3 \times 10^{5}$ in Figures \ref{fig:3} and \ref{fig:4}, respectively, for the three different wavebands and three different PSD models of the Lorentz factor distribution ($\nu_B$ $=$ $10^{-5}$, $10^{-6}$ and $10^{-7}$ Hz). For all of the above parameter space, multiple shocks are introduced in the emission region at injection times generated by the collision of successive blobs as described in the previous section. Therefore, the large flaring events in the light curves are coincident with the shock injection times. In Figures \ref{fig:3a} and \ref{fig:4a}, we show the light curves for the different shock amplitudes scenario, and in Figures \ref{fig:3b} and \ref{fig:4b}, for the case, in which all shocks are of the same amplitude.  We can see some large and prominent outbursts in Figures \ref{fig:3a} and \ref{fig:4a}, along with many small flaring events. This is due to the fact that we have considered different amplitudes of the shocks in this case and the amplitudes of some of the shocks, as calculated in our simulation, are large. On the other hand, in the same shock amplitude scenario, we can see in Figures \ref{fig:3b} and \ref{fig:4b} that the resultant flares are approximately of similar amplitude, with very large or very small flares being rare. For a quantitative comparison, we calculated the mean $\mu$ and the standard deviation $\sigma$ for our light curves in both the scenarios, and found that the peak of a larger fraction of outbursts are limited within $\mu$ and $\mu + 2\sigma$ for the same shock amplitude scenario than that in the different shock amplitude case.

From Figures \ref{fig:3} and \ref{fig:4}, we note that for $\gamma_{max}$ $=$ $3 \times 10^{3}$, the flares in the $\gamma$-ray light curves decay faster than their X-ray counterparts. While emission at both of those wavebands are due to the IC process, the former is produced by higher energy electrons, which have faster radiative cooling. On the other hand, for $\gamma_{max}$ $=$ $3 \times 10^{5}$, the $\gamma$-ray flares decay more slowly than those at X-rays because the former is due to the IC process by lower energy electrons while the X-ray flares, which arise from synchrotron radiation, are generated by higher energy electrons near the peak of the electron energy distribution. This is also apparent from the infrared and $\gamma$-ray light curves, since the flares in those wavebands decay almost at a similar rate. Both of them arise from electrons having similar energy, with the former being generated by synchrotron emission and the latter by IC. The above arguments also explains the slower and faster decay of X-ray flares in comparison to the infrared ones, for $\gamma_{max}$ $=$ $3 \times 10^{3}$ and $\gamma_{max}$ $=$ $3 \times 10^{5}$, respectively. Our results, obtained using a very simple version of the internal shocks model, are in agreement with the ones using a more detailed version of it \citep{Boettcher&Dermer2010, Joshi&Boettcher2011}. For example, \cite{Joshi&Boettcher2011} showed that their X-ray light curves, which originated due to the up-scattering of lower energy synchrotron photons off lower energy electrons, have a longer decay timescale. But their optical and $\gamma$-ray light curves have a shorter decay timescale, since the former is produced by the synchrotron process and the latter is produced by the IC scattering by a higher energy electron distribution in both cases.

\subsection{Flux Distribution}
\label{sec: flux dist}

\begin{figure*}
    \begin{subfigure}{0.49\textwidth}
        \centering
        \includegraphics[width=\textwidth, height=2\textwidth]{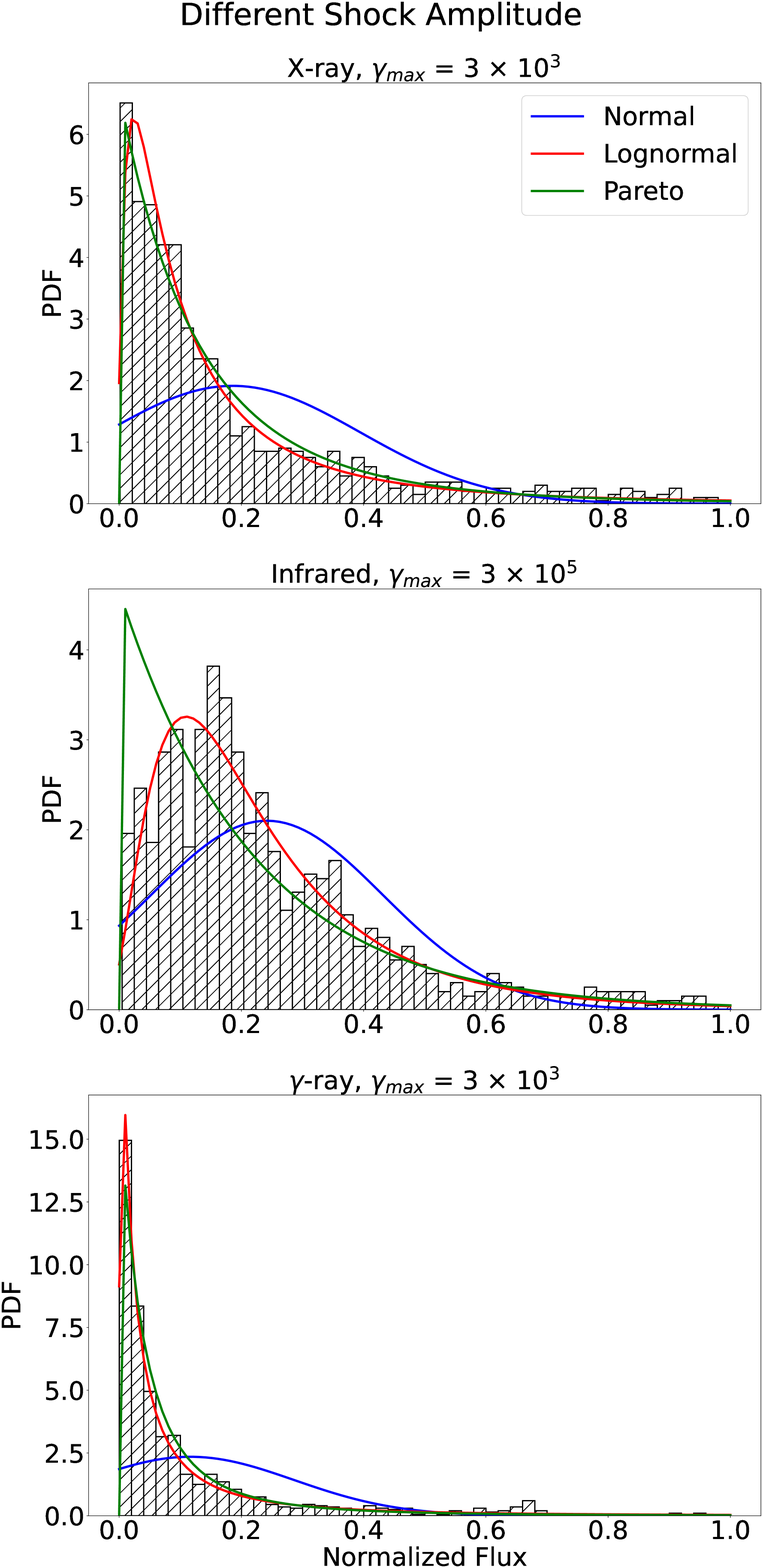}
        \caption{}
        \label{fig:5a}
    \end{subfigure}
    \hfill%
    \begin{subfigure}{0.49\textwidth}
        \centering
        \includegraphics[width=\textwidth, height=2\textwidth]{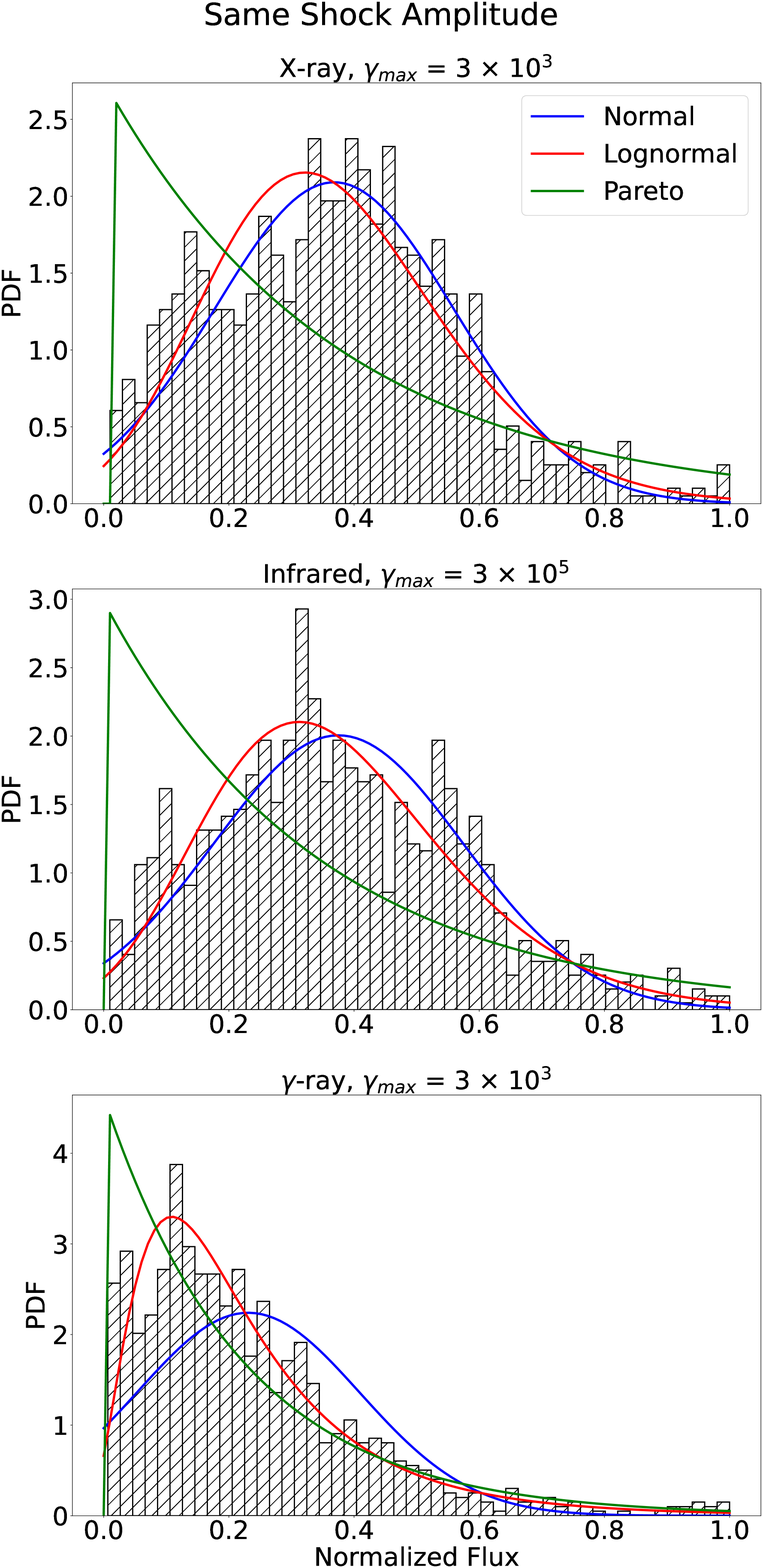}
        \caption{}
        \label{fig:5b}
    \end{subfigure}
    \caption{Black hashed histograms denote the flux distribution of the light curves in (a) different and (b) same shock amplitude scenario, respectively. Blue, red, and green solid lines indicate the best-fit Normal, Log-normal and Pareto distributions, respectively. Different shock amplitude scenario gives rise to a more skewed flux distribution in comparison to the same shock amplitude scenario, and faster decaying light curves exhibit a more skewed distribution.}
    \label{fig:5}
\end{figure*}

The flux distribution of a long-term time series provides a statistical description of the underlying stochastic processes that drive the variability. In an additive process, in which the fluctuations are drawn randomly and are independent of each other, the flux values follow a Gaussian distribution according to the central limit theorem \citep{Uttley&etal2005}. On the other hand, a multiplicative process gives rise to a skewed distribution with a high-end flux tail. As found in BHXRB Cyg X-1 \citep{Uttley&etal2005}, a protostar IRAS 13224–3809 \citep{Alston&etal2019} and most recently in blazars \citep{Kushwaha&etal2017, Shah&etal2018, Sinha&etal2018, Khatoon&etal2020}, log-normal flux distribution is the most common one. A skewed distribution, most commonly log-normal, is often associated with a linear rms-flux relation. We studied the flux distribution of our simulated light curves and attempted to fit them with normal, log-normal and Pareto distribution, while using the Kolmogorov-Smirnov (KS) test to determine the best possible fit. We show the flux distribution of some of our simulated light curves for the bulk Lorentz factor PSD model with break frequency $\nu_B$ $=$ $10^{-6}$ Hz, in both the different and same shock amplitude scenario in Fig. \ref{fig:5}. We also show the results of the KS test in Table \ref{table:2}. We find that the light curves in the different shock amplitude scenario produce highly skewed flux distributions compared to those in the same shock amplitude case. This is because in the same amplitude scenario, a larger fraction of higher flux values may be obtained. In general, we find that the light curves in which the flares decay faster, produce a more skewed distribution in comparison to a slowly decaying light curve, since there is a smaller chance of overlap of flares and hence a greater probability of lower flux values.

\begin{table*}
    \centering
    \caption{Results of the Kolmogorov-Smirnov (KS) test performed on the flux distributions (PSD break frequency of the Lorentz factor distribution, $\nu_B$ $=$ $10^{-6}$ Hz).}
    \setlength{\tabcolsep}{22pt}
    \renewcommand{\arraystretch}{2}
    
    \begin{tabular}{ccccccc}
        \hline\hline
        \multicolumn{5}{c}{Different shock amplitude scenario}\\
        \hline
        Light curve & Skewness & \makecell{Distribution} & $D$-statistic & $p$-value \\ 
        \hline\hline

        \multirow{3}{*}{X-ray; $\gamma_{max}$ $=$ $3 \times 10^3$} & \multirow{3}{*}{1.783} & Log-normal & 0.028 & 0.41 \\
        
        & & Pareto & 0.033 & 0.214 \\
        
        & & Normal & 0.194 & < 0.001 \\
        
        \hline
        
        \multirow{3}{*}{Infrared; $\gamma_{max}$ $=$ $3 \times 10^5$} & \multirow{3}{*}{1.488} & Log-normal & 0.031 & 0.296 \\
        
        & & Pareto & 0.141 & < 0.001 \\
        
        & & Normal & 0.129 & < 0.001 \\
        
        \hline
        
        \multirow{3}{*}{$\gamma$-ray; $\gamma_{max}$ $=$ $3 \times 10^3$} & \multirow{3}{*}{2.159} & Log-normal & 0.034 & 0.182 \\
        
        & & Pareto & 0.051 & 0.01 \\ 
        
        & & Normal & 0.247 & < 0.001 \\
        
        \hline
        
        \multicolumn{5}{c}{Same shock amplitude scenario}\\
        \hline
        \makecell{Light curve} & \makecell{Skewness} & \makecell{Distribution} & $D$-statistic & \makecell{$p$-value} \\ 
        \hline\hline
        
        \multirow{3}{*}{X-ray; $\gamma_{max}$ $=$ $3 \times 10^3$} & \multirow{3}{*}{0.398} & Log-normal & 0.041 & 0.062 \\
        
        & & Pareto & 0.191 & < 0.001 \\
        
        & & Normal & 0.036 & 0.14 \\
        
        \hline
        
        \multirow{3}{*}{Infrared; $\gamma_{max}$ $=$ $3 \times 10^5$} & \multirow{3}{*}{0.46} & Log-normal & 0.036 & 0.148 \\
        
        & & Pareto & 0.226 & < 0.001 \\
        
        & & Normal & 0.049 &  0.015 \\

        \hline
        
        \multirow{3}{*}{$\gamma$-ray; $\gamma_{max}$ $=$ $3 \times 10^3$} & \multirow{3}{*}{1.472} & Log-normal & 0.032 & 0.251 \\
        
        & & Pareto & 0.116 & < 0.001 \\
        
        & & Normal & 0.104 & < 0.001 \\

        \hline
        
    \end{tabular}
    
    \label{table:2}
    
\end{table*}

\subsection{RMS-Flux relation}

\begin{table*}
    \centering
    \caption{Results related to rms-flux relation for the different shock amplitude scenario.}
    \setlength{\tabcolsep}{16pt}
    \renewcommand{\arraystretch}{2.5}
    
    \begin{tabular}{cccccccc}
        \hline\hline
        \multicolumn{6}{c}{$\gamma_{max}$ $=$ $3 \times 10^{3}$ ; $\nu_1$ to $\nu_2$ $=$ $ 4 \times 10^{-7}$ to $4 \times 10^{-6}$ Hz}\\
        \hline
        \makecell{$\nu_B$ (LF) \\(in Hz)} & \makecell{Frequency \\ (in Hz)} & \makecell{$T_{seg}$ \\ (in days)} & $F_{bin}$ & \makecell{Slope \\ $m$} & \makecell{Intercept \\ $c$} \\
        \hline\hline

        \multirow{3}{*}{$10^{-5}$} & Infrared; $10^{12.0}$  & 58 & 5 & \makecell{0.269\\$\pm$ 0.010} & \makecell{0.000\\$\pm$ 0.001}\\
        
        & X-ray; $10^{18.0}$  & 56 & 5 & \makecell{0.230\\$\pm$0.015} & \makecell{0.010\\$\pm$ 0.003}\\
        
        & Gamma; $10^{20.6}$  & 63 & 5 & \makecell{0.281\\$\pm$ 0.018} & \makecell{0.007\\$\pm$ 0.003}\\
        \hline

        \multirow{3}{*}{$10^{-6}$} & Infrared; $10^{12.0}$  & 69 & 5 & \makecell{0.253\\$\pm$ 0.012} & \makecell{0.015\\$\pm$0.003}\\
        
        & X-ray; $10^{18.0}$  & 57 & 5 & \makecell{0.237\\$\pm$ 0.006} & \makecell{0.002\\$\pm$0.001}\\
        
        & Gamma; $10^{20.6}$  & 64 & 5 & \makecell{0.277\\$\pm$0.010} & \makecell{0.009\\$\pm$0.001}\\
        \hline

        \multirow{3}{*}{$10^{-7}$} & Infrared; $10^{12.0}$  & 64 & 4 & \makecell{0.222\\$\pm$ 0.014} & \makecell{0.018\\$\pm$0.003}\\
        
        & X-ray; $10^{18.0}$  & 53 & 5 & \makecell{0.202\\$\pm$0.006} & \makecell{0.005\\$\pm$0.001}\\
        
        & Gamma; $10^{20.6}$  & 66 & 4 & \makecell{0.319\\$\pm$0.011} & \makecell{0.004\\$\pm$0.001}\\
        \hline

        \multicolumn{6}{c}{$\gamma_{max}$ $=$ $3 \times 10^{5}$ ; $\nu_1$ to $\nu_2$ $=$ $4 \times 10^{-7}$ to $4 \times 10^{-6}$ Hz}\\
        \hline
        \makecell{$\nu_B$ (LF) \\(in Hz)} & \makecell{Frequency \\ (in Hz)} & \makecell{$T_{seg}$ \\ (in days)} & $F_{bin}$ & \makecell{Slope \\ $m$} & \makecell{Intercept \\ $c$} \\
        \hline\hline
        \multirow{3}{*}{$10^{-5}$} & Infrared; $10^{12.0}$  & 63 & 5 & \makecell{0.278\\$\pm$0.021} & \makecell{0.009\\$\pm$ 0.004}\\
        
        & X-ray; $10^{18.0}$  & 58 & 5 & \makecell{0.466\\$\pm$ 0.011} & \makecell{0.011\\$\pm$0.001}\\
        
        & Gamma; $10^{20.6}$  & 63 & 5 & \makecell{0.227\\$\pm$0.017} & \makecell{0.012\\$\pm$0.003}\\

        \hline
        \multirow{3}{*}{$10^{-6}$} & Infrared; $10^{12.0}$  & 62 & 4 & \makecell{0.326\\$\pm$ 0.017} & \makecell{0.002\\$\pm$0.004}\\
        
        & X-ray; $10^{18.0}$  & 68 & 5 & \makecell{0.613\\$\pm$0.011} & \makecell{0.001\\$\pm$0.001}\\
        
        & Gamma; $10^{20.6}$  & 64 & 4 & \makecell{0.251\\$\pm$0.012} & \makecell{0.011\\$\pm$0.001}\\

        \hline
        \multirow{3}{*}{$10^{-7}$} & Infrared; $10^{12.0}$  & 64 & 5 & \makecell{0.237\\$\pm$0.013} & \makecell{0.014\\$\pm$0.003}\\
        
        & X-ray; $10^{18.0}$  & 55 & 4 & \makecell{0.918\\$\pm$0.021} & \makecell{0.000\\$\pm$0.001}\\
        
        & Gamma; $10^{20.6}$  & 54 & 4 & \makecell{0.274\\$\pm$0.009} & \makecell{0.002\\$\pm$0.001}\\
        \hline
        
    \end{tabular}
    
    \label{table:3}
    
\end{table*}

We follow the prescription of \cite{Heil&etal2012} to check for the rms-flux relation in our simulated light curves. We first divide the time-series into a certain number of time segments. For each segment $i$, we calculate the mean flux $\braket{x_i}$ and the power $P_i(\nu_j)$ for a particular frequency of variation $\nu_j$. We estimate the power spectrum using the absolute normalization, so that the variance is equal to the integrated area under the power spectrum. Due to the stochastic nature of the light curves, the estimated power values are scattered around the actual value of power according to a $\chi^2$ distribution with two degrees of freedom \citep{Heil&etal2012}. So to reduce this effect we arrange $\braket{x_i}$ into a number of flux bins and take the average of the mean flux and the power $P_i(\nu_j)$ in each of these bins as follows

\begin{subequations}

    \begin{equation}
        \overline{x} = \frac{1}{M} \sum_{k=1}^M \braket{x_k}
        \label{eqn: 12a}
    \end{equation}

    \begin{equation}
        \overline{P(\nu_j)} = \frac{1}{M} \sum_{k=1}^M P_k(\nu_j),
        \label{eqn: 12b}
    \end{equation}
\end{subequations}

where $M$ is the number of power spectrum values in each flux bin and $\overline{x}$ is the final mean flux for that bin.

\medbreak
The variance corresponding to each of these mean fluxes is calculated by integrating the average power spectrum over a frequency range $\nu_1$ to $\nu_2$ as follows, 

\begin{equation}
    \sigma^2 = \sum_{j=1}^{W}\overline{P(\nu_j)}\Delta \nu,
    \label{eqn: 13}
\end{equation}

where $W$ is the number of power spectrum points in the chosen frequency range, $\Delta \nu$ $=$ $1/T$, is the frequency resolution of the power spectrum, and $T$ is the total length of the time series.

\medbreak
By taking the square root of the variance we get the rms value. The uncertainty on the rms is given by the formula \citep{Heil&etal2012},

\begin{equation}
    sd[\sigma] \approx \frac{\Delta \nu}{2 \sigma M^{1/2}} \Bigg( \sum_{j=1}^W \overline{P^2(\nu_j)} \Bigg)^{1/2}.
    \label{eqn: 14}
\end{equation}

\begin{figure*}
    \begin{subfigure}{0.49\textwidth}
        \centering
        \includegraphics[width=\textwidth, height=2\textwidth]{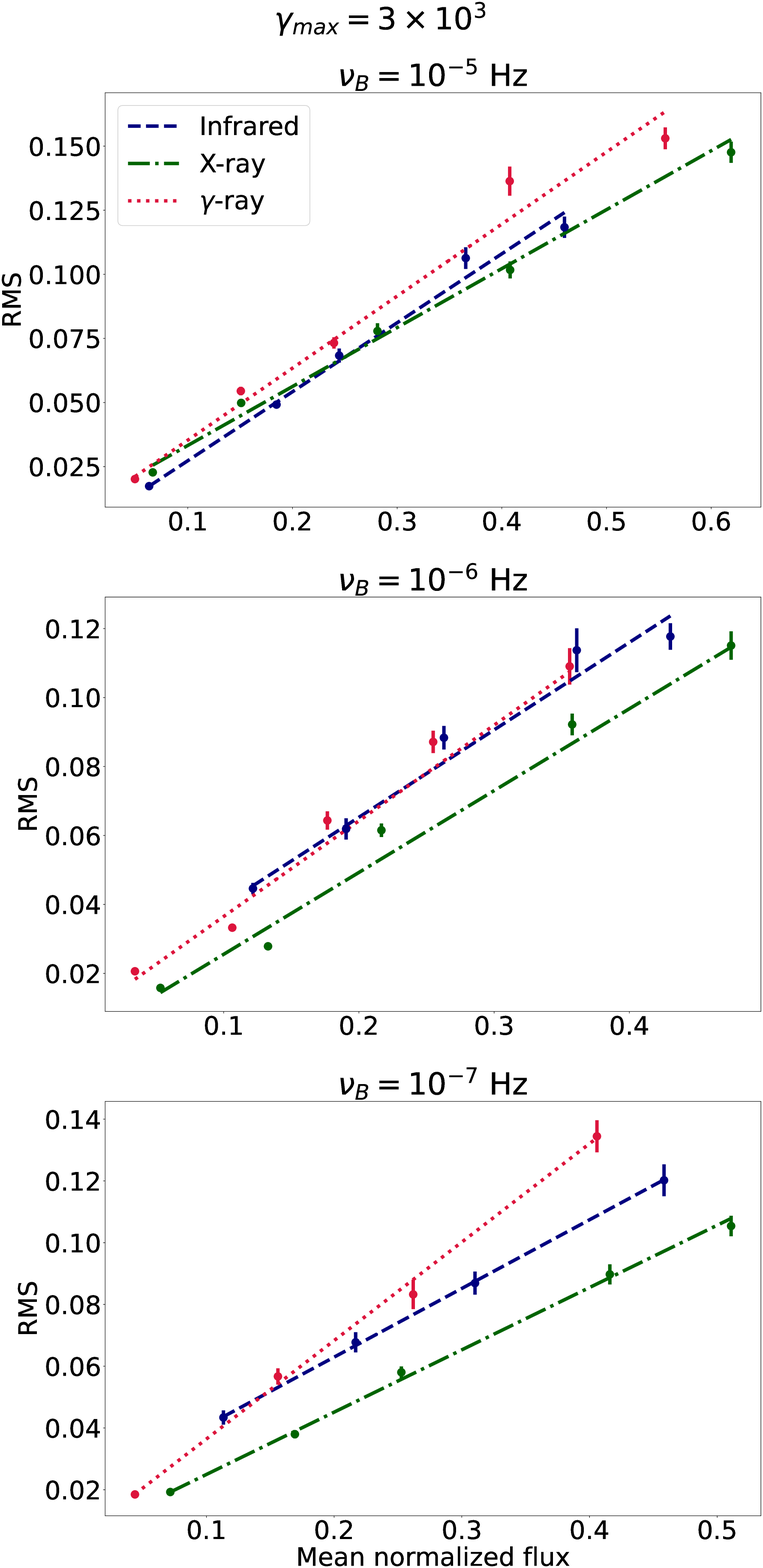}
        \caption{}
        \label{fig:6a}
    \end{subfigure}
    \hfill%
    \begin{subfigure}{0.49\textwidth}
        \centering
        \includegraphics[width=\textwidth, height=2\textwidth]{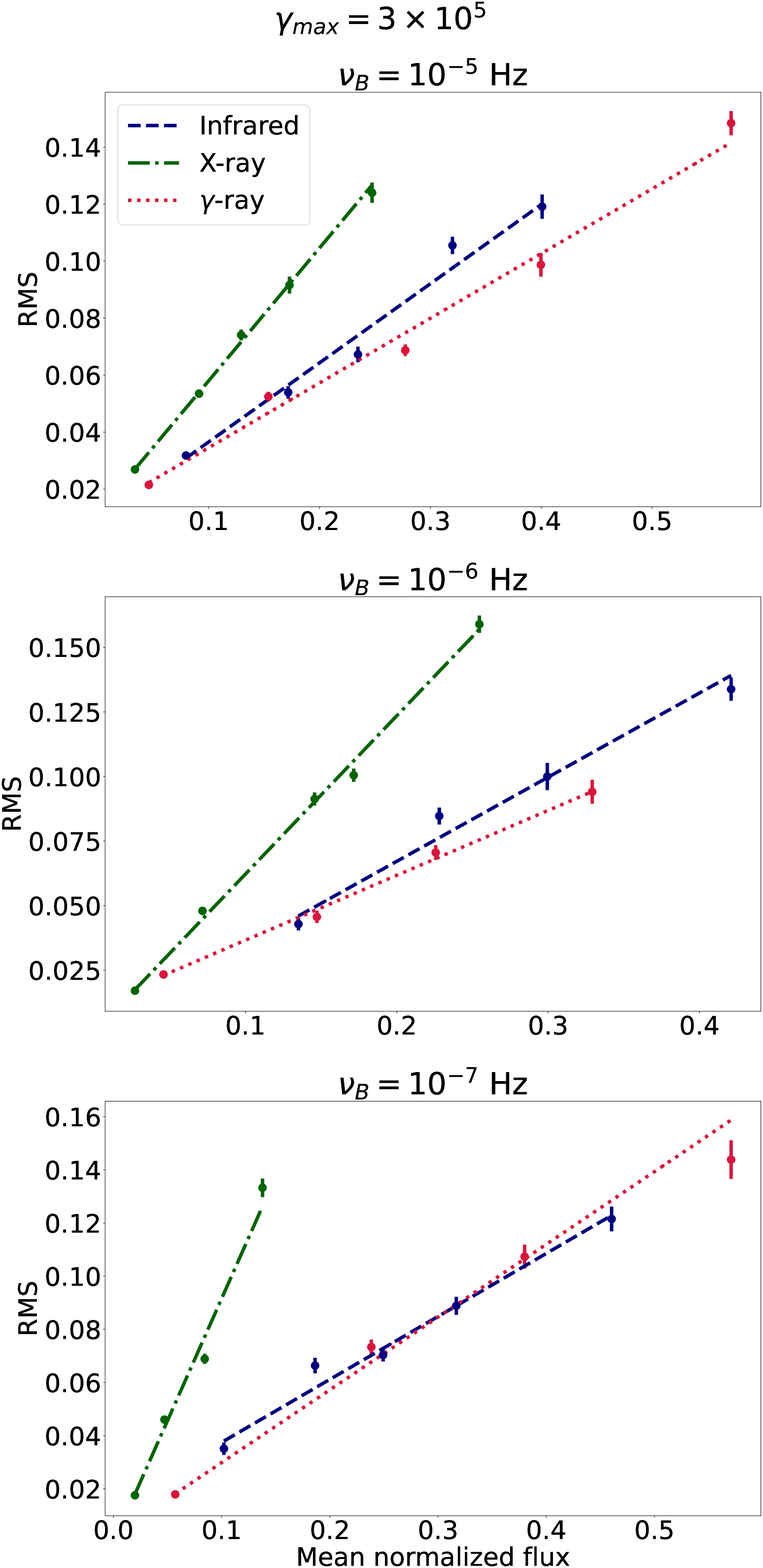}
        \caption{}
        \label{fig:6b}
    \end{subfigure}
    \caption{RMS-flux relation for (a) $\gamma_{max}$ $=$ $3 \times 10^{3}$ and (b) $\gamma_{max}$ $=$ $3 \times 10^{5}$, in the different shock amplitude scenario. Blue, green, and red filled circles with error bars denote the infrared ($10^{12.0}$ Hz), X-ray ($10^{18}$ Hz), and $\gamma$-ray ($10^{20.6}$ Hz) wavebands, respectively. Blue dashed, green dot-dashed and red dotted lines represent the best-fit straight lines through the respective points. The three rows represent the three different break-frequencies: $10^{-5}$ Hz, $10^{-6}$ Hz and $10^{-7}$ Hz of the PSD of the Lorentz factor distribution. The slope is found to be dependent on the wavelength of the light curve as well as on $\gamma_{max}$, i.e., the electron population.}
    \label{fig:6}
\end{figure*}

\begin{figure*}
    \begin{subfigure}{0.49\textwidth}
        \centering
        \includegraphics[width=\textwidth, height=0.8\textwidth]{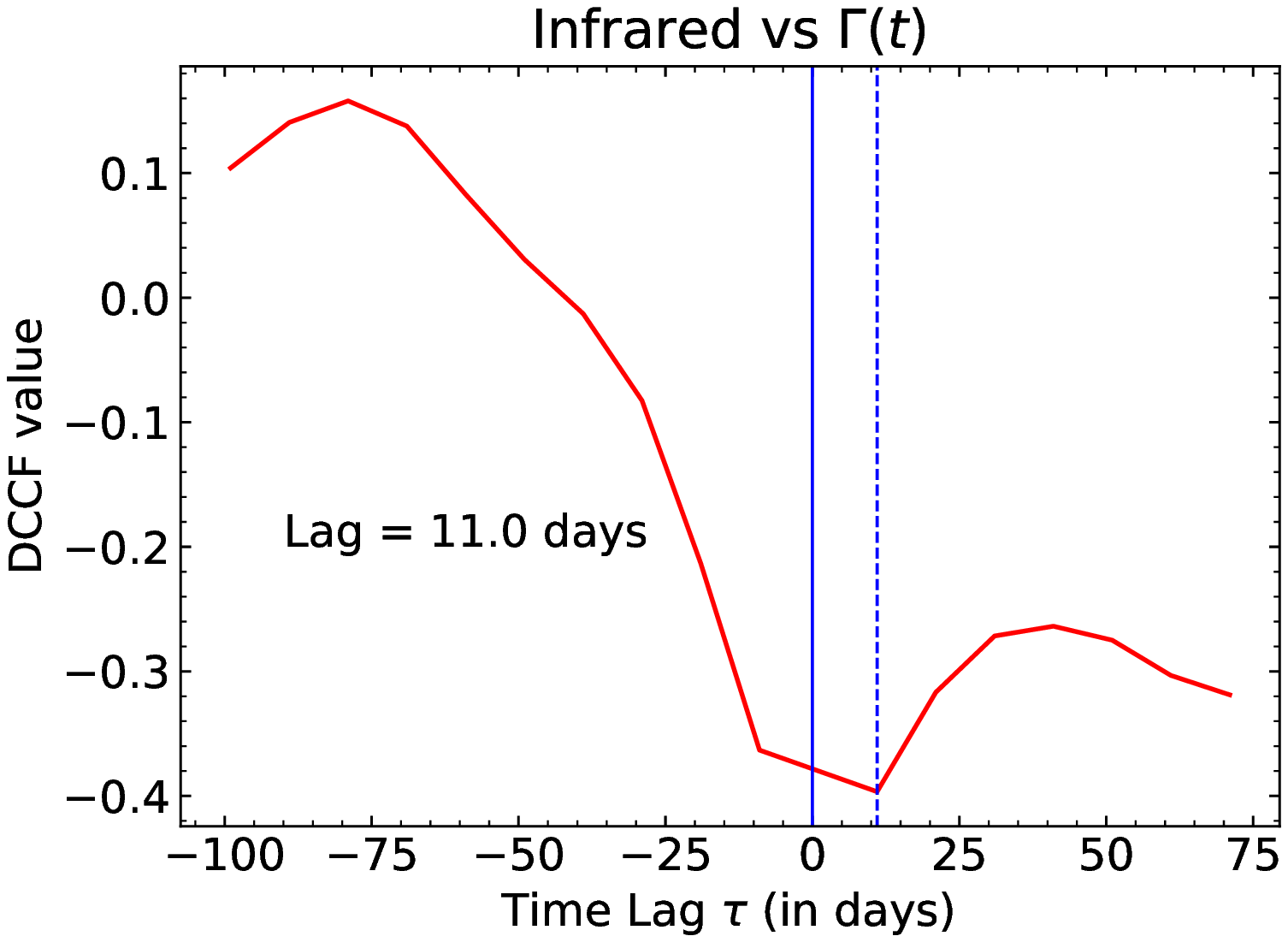}
        \caption{}
        \label{fig:7a}
    \end{subfigure}
    \hfill%
    \begin{subfigure}{0.49\textwidth}
        \centering
        \includegraphics[width=\textwidth, height=0.8\textwidth]{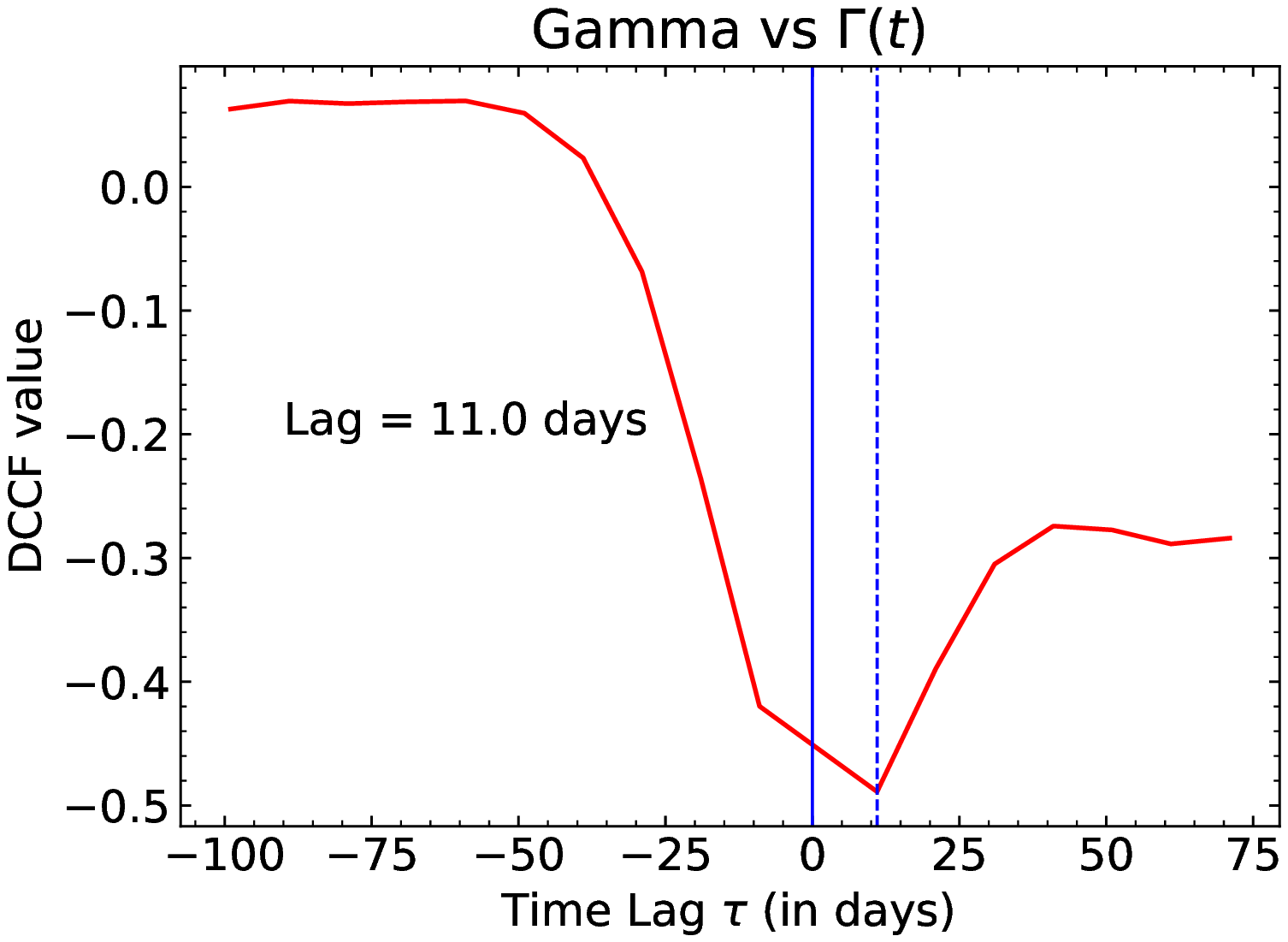}
        \caption{}
        \label{fig:7b}
    \end{subfigure}
    \caption{The discrete cross-correlation function between jet (a) infrared ($10^{12}$ Hz), (b) $\gamma$-ray ($10^{20.6}$ Hz)) and disc light curves. The solid line denotes the zero time lag while the dotted line indicates the time lag corresponding to the maximum DCCF value. The infrared and $\gamma$-ray variability both lag that of the disc by $\sim$11 days, respectively. Here $\gamma_{max}$ $=$ $3 \times 10^3$.}
    \label{fig:7}
\end{figure*}

\medbreak
We check for a linear rms-flux relation in each of the simulated light curves by using a function $\sigma$ $=$ $m \overline{x} + c$ and estimate the best-fit slope $m$ and intercept $c$. We calculate the coefficient of determination, $r^2$ value, for each of the fit and put a stringent value of $r^2$ $=$ $0.95$ as our criteria for a good rms-flux relation. In Table \ref{table:3}, we show the best-fit parameters of the linear relation, along with the length of the time segments ($T_{seg}$), the number of flux bins ($F_{bin}$) and the frequency range used for integration ($\nu_1$ to $\nu_2$) in each of the light curves for different shock amplitude scenario. We chose the frequency range of integration from $4  \times 10^{-7}$ Hz to $4 \times 10^{-6}$ Hz. We use the conventional way of choosing the frequency range for integration to be one decade \citep{Uttley&McHardy2001, Gleissner&etal2004, Uttley&etal2005, Gandhi2009, Heil&etal2012} and our chosen values approximately span the middle region of the total frequency range corresponding to the light curve segments so that most of the variability power is incorporated without including the low- and high-frequency ends, which may contain unwanted noise. In Figures \ref{fig:6a} and \ref{fig:6b}, we show the rms-flux relations for $\gamma_{max}$ $=$ $3 \times 10^3$ and $3 \times 10^5$, respectively, for the different shock amplitude scenario.

We study the multiplicative nature of blazar variability by testing if the simulated light curves exhibit the rms-flux relation because the latter has been shown to be a property of a multiplicative time series \citep{Uttley&etal2005, Vaughan&etal2003}. We can see in Table \ref{table:3} that in the different shock amplitude scenario, we get a linear rms-flux relation in all the light curves for both the cases of $\gamma_{max}$. We chose different length of the time segments, ranging from $40$ to $90$ days in steps of $1$ day. This range was chosen to optimize the number of points in each segment and the total number of segments. We obtained a linear relation for $\sim$26-32\% of our choices of segment lengths across the three wavelengths and in each case of the bulk Lorentz factor break frequency. However, in the same shock amplitude scenario, for $\gamma_{max}$ $=$ $3 \times 10^3$, we obtain a linear rms-flux relation for $\sim$14\%, $\sim$6\%, and $\sim$18\% of our choices in the case of infrared, X-ray and $\gamma$-ray, respectively. For $\gamma_{max}$ $=$ $3 \times 10^5$, while $\sim$12\% and $\sim$14\% of our choices in the case of infrared and $\gamma$-ray, respectively, exhibited a linear rms-flux relation, for X-rays the fraction is $\sim$22\%. When we lowered the constraint of $r^2$ value from 0.95 to 0.5, our percentage of getting an rms-flux relation in the different shock amplitude scenario became $\sim$85-95\%, while in the same shock amplitude scenario it rose to $\sim$45-55\% in the case of infrared and $\gamma$-ray for both the $\gamma_{max}$ values. In the case of X-ray, the percentage became $\sim$50\% for the lower $\gamma_{max}$ value and $\sim$93\% for the higher $\gamma_{max}$ value. By considering the consistently high percentage of time segments that produce a linear rms-flux relation in the different shock amplitude scenario, we argue that it imparts a multiplicative nature in the light curves, across all the wavelengths. On the other hand, in the same shock amplitude scenario, the occurrence of a linear rms-flux relation depends on the wavelength as well as on the value of $\gamma_{max}$. This is consistent with what we found in Section \ref{sec: flux dist} that even in the same shock amplitude scenario for some values of $\gamma_{max}$ and considered wavelength of the light curve the flux distribution is skewed, and the findings in literature that a linear rms-flux relation is obtained if the flux distribution is log-normal or highly skewed \citep{Uttley&etal2005, Alston&etal2019, Biteau&Giebels2012}.

In Table \ref{table:3}, we can see that for $\gamma_{max}$ $=$ $3 \times 10^{3}$, the rms-flux relation for all the three wavelengths have similar slopes, with slightly smaller values in the case of X-ray. The slope is related to the fractional rms (rms normalized by mean flux), which in turn, depends on the state of the source. Usually in the case of X-ray binaries, soft states are associated with lower slope values than the hard states \citep{Gleissner&etal2004, Heil&etal2012}. Here we put forth an argument to explain the dependence of the slope on the wavelength of the light curves as well as on the maximum energy of the electron distribution. We find that the slope of the linear rms-flux relation obtained from a given light curve is higher if the decay timescale of the flares in it is shorter. If the flares decay faster then the light curve contains fewer overlapping flares, which in turn, causes the mean flux in time segments to be lower. That results in a higher value of the slope. For example, in Figure \ref{fig:3a}, the X-ray flares decay slightly more slowly, i.e., are wider than the $\gamma$-ray light curves. As a result, the slope is slightly smaller in the X-rays.  
In Figure \ref{fig:3b} ($\gamma_{max}$ $=$ $3 \times 10^{5}$), the X-ray flares decay faster than their low $\gamma_{max}$ counterparts causing a higher value of the slope of the rms-flux relation compared to that in the case of $\gamma_{max}$ $=$ $3 \times 10^{3}$. The low value of the slope in the case of $\gamma$-rays and infrared in Figure \ref{fig:6b}, when compared to that in the X-rays, is also due to the slower decay of the $\gamma$-ray and infrared flares than their X-ray counterparts because in that case X-ray emission is produced by higher energy electrons through synchrotron radiation while the $\gamma$-rays and infrared emission are produced by relatively lower energy electrons through the IC and synchrotron processes, respectively.

\subsection{Correlation between disc and jet light curves}

We have followed \cite{Malzac&etal2018}, by assuming that the Lorentz factor fluctuations trace the disc variability. They fitted the SED of the BHXRB GX 339-4 using the internal shocks model and also analysed various temporal characteristics by assuming different dependencies of the Lorentz factor fluctuations on the disc variability. Upon assuming constant mass of the blobs, they showed that the linear dependency can provide better agreement to the data. To find any temporal correlation between the disc and jet variability, we have calculated the discrete cross-correlation function \cite[DCCF;][]{Edelson&Krolik1988} between the jet and the disc light curve, in which the latter is represented by the Lorentz factor fluctuations.

We smoothed the light curves using a Gaussian kernel of standard deviation $10$ days before calculating the DCCF. This was done in order to avoid the effect of small scale fluctuations in the disc and jet light curves and to probe the correlation by considering only the large scale fluctuations which arise due to the injection of the shocks. Fig. \ref{fig:7} shows the DCCF between the disc and jet variability, at infrared and $\gamma$-ray wavebands, respectively, for $\nu_B$ $=$ $10^{-6}$ Hz and $\gamma_{max}$ $=$ $3 \times 10^3$. We see an anti-correlation between the jet and disc light curves, with the infrared and $\gamma$-ray variability, both lagging behind the disc light curves by $\sim$11 days. We have also checked the cross-correlation function in other wavelengths and for different values of $\nu_B$ and $\gamma_{max}$ and found similar results as above. A dip in the Lorentz factor distribution implies the generation of a faster blob following a slower one, which would lead to a collision down the jet, which eventually results in the energising of electrons and a flare in the jet emission. Based on our assumption that the Lorentz factor series depends linearly on the variable disc emission, this connection may manifest as a delayed anti-correlation of disc-jet emission variability, as we are seeing here.

\section{Summary and Conclusions}

In this work, we have explored the physical mechanisms responsible for the rms-flux relation in blazar variability in the context of the internal shocks model. We have performed a detailed simulation of emission from blazars by incorporating the non-thermal processes in the jet, to probe the disc-jet connection in blazars via this new window. We summarize our findings below: 

\begin{enumerate}

    \item \indent We verify that the simulated SED exhibits the double-hump nature of observed blazar SEDs and the synchrotron peak frequency increases with the increase in the maximum Lorentz factor of electrons, as expected. In addition, we find that our simulated light curves successfully reproduce the high amplitude variability as seen in observed blazar light curves.

    \item The emergence of a linear rms-flux relation in the simulated blazar light curves points to an inherent multiplicative nature of the underlying stochastic processes. We show that different amplitudes of the injected shocks can consistently produce a linear rms-flux relation across all the wavelengths, while the same shock amplitude scenario is inconsistent in producing an rms-flux relation. The occurrence of the rms-flux relation is found to be dependent on the wavelength as well on the electron population through $\gamma_{max}$. This can be linked to the decay timescale of the flares, on which the flux distribution depends, since we found that faster decaying flares produce a flux distribution with more skewness, which in turn, produces a linear rms-flux relation.
    
    \item We find that the slope of the linear rms-flux relation obtained from a given light curve is higher if the decay timescale of the flares in it is shorter. If the flares decay faster then the light curve contains fewer overlapping flares, which in turn, causes the mean flux in time segments to be lower. That results in a higher value of the slope.

    \item Assuming the Lorentz factor distribution to be approximately proportional to the disc light curve, we find that the jet variability is anti-correlated with that of the disc with the former lagging that of the latter by $\sim$11 days. This is similar to what has been observed in some X-ray binaries and AGN. 

\medbreak    
We have found that blobs of plasma ejected at regular intervals and having equal mass but different speeds may collide while travelling down the jet and if the resultant shock fronts are assumed to energize the electrons in the jet, which subsequently cool through non-thermal processes then we can reproduce many of the observed properties of multi-wavelength blazar variability. The reproduction of the linear rms-flux relation and the disc-jet delayed anti-correlation in our light curves simulated under certain conditions in the broad context of the internal shocks model provide a possible mode of connection between the disc and jet in AGN, which may be verified with current or future observation of the dynamics of the pc-scale jet. Furthermore, our analyses make clear predictions about the nature of the rms-flux relation at different wavebands and for blazars with two ranges of values of the peak synchrotron frequency.

\end{enumerate}

\section*{Acknowledgements}

We thank the anonymous referee whose comments and suggestions have helped to improve the manuscript. We are grateful to Phil Uttley and Souradip Bhattacharyya for useful discussions. RC thanks Presidency University for support under the Faculty Research and Professional Development (FRPDF) Grant, ISRO for support under the AstroSat archival data utilization program, and IUCAA for their hospitality and usage of their facilities during his stay at different times as part of the university associateship program. RC acknowledges financial support from BRNS through a project grant (sanction no: 57/14/10/2019-BRNS) and thanks the project coordinator Pratik Majumdar for support regarding the BRNS project. 

\section*{Data Availability}

The simulated data underlying this article will be shared on reasonable request to the corresponding author.



\bibliographystyle{mnras}
\bibliography{references} 








\bsp	
\label{lastpage}
\end{document}